\documentclass[11pt, oneside]{article}   	% use "amsart" instead of "article" for AMSLaTeX format
\usepackage{amsmath, amsthm, amssymb}
\usepackage{tabu}
\usepackage{caption}
\usepackage{array}
\usepackage{geometry}                		% See geometry.pdf to learn the layout options. There are lots.
\usepackage{mathtools}
\usepackage{morefloats}
\usepackage[parfill]{parskip}    		% Activate to begin paragraphs with an empty line rather than an indent
\usepackage{graphicx}				% Use pdf, png, jpg, or eps with pdflatex; use eps in DVI mode
\usepackage{caption}								% TeX will automatically convert eps --> pdf in pdflatex	

\usepackage{enumitem}												
\usepackage{amssymb}

\title{Light Trapping Textures Designed by Electromagnetic Optimization for Sub-Wavelength Thick Solar Cells}
\author{Vidya Ganapati, Owen D. Miller, Eli Yablonovitch \\ \\University of California, Berkeley \\ Material Sciences Division, Lawrence Berkeley National Laboratory}
\begin{document}
\maketitle
\thispagestyle{empty}
\pagestyle{empty}
%\section{}
%\subsection{}

%$$ $$ - separate line equation
%$ $

%{} \frac {}

\section*{Abstract}

Light trapping in solar cells allows for increased current and voltage, as well as reduced materials cost. It is known that in geometrical optics, a maximum $4n^2$ absorption enhancement factor can be achieved by randomly texturing the surface of the solar cell, where $n$ is the material refractive index. This ray-optics absorption enhancement limit only holds when the thickness of the solar cell is much greater than the optical wavelength. In sub-wavelength thin films, the fundamental questions remain unanswered: 
\begin{enumerate}[label={(\arabic*)}]
\item what is the sub-wavelength absorption enhancement limit and 
\item what surface texture realizes this optimal absorption enhancement? 
\end{enumerate}
We turn to computational electromagnetic optimization in order to design nanoscale textures for light trapping in sub-wavelength thin films. For high-index thin films, in the weakly absorbing limit, our optimized surface textures yield an angle- and frequency-averaged enhancement factor $\approx 39$. They perform roughly 30\% better than randomly textured structures, but they fall short of the ray optics enhancement limit of $4n^2 \approx 50$.

\section*{Introduction} 

Texturing of solar cell surfaces allows for absorption enhancement, owing to the coupling of incident light rays to totally internally reflected modes within the cell, i.e. light trapping. It is known that in the ray-optics regime, where the thickness of the solar cell is much greater than the wavelength of light, that the maximum absorption for weakly absorbed rays is given by \cite{yablonovitch_statistical_1982}: 
\begin{equation}\label{eq:rayOpticsA} A = \frac{\alpha d}{\alpha d + \frac{1}{4 n^2}},\end{equation}
where $\alpha$ is the absorption coefficient, $d$ the thickness of the material, and $n$ the index of refraction. This maximum absorption limit assumes a perfect rear mirror. We can compare this to the single-pass absorption of weakly absorbed light:
\begin{equation}\label{eq:singlePass}A = 1 - e^{-\alpha d} \approx \alpha d. \end{equation}
The absorption enhancement is the actual absorption divided by the single pass absorption. The maximum absorption enhancement in the ray-optics regime is thus given by Eqn.~\ref{eq:rayOpticsA} divided by Eqn.~\ref{eq:singlePass}; in the limit of a very weakly absorbing material the absorption enhancement is given by $4n^2$.

	With light trapping, we can achieve high absorption, even for thin absorber layers.  Short circuit current ($J_{SC}$) and fill factor improvements occur due to better carrier extraction in thin layers. Additionally, open circuit voltage ($V_{OC}$) improvements occur, owing to increased carrier concentration. In high quality materials, such as gallium arsenide, efficiency improvement can be substantial, due to improvement in external fluorescence yield \cite{miller_strong_2012, lush_thin_1991}. We also reduce material cost by achieving the same current in a thinner material.
	
	In recent years, light trapping has seen renewed interest in the sub-wavelength regime, which is applicable to increasingly thin solar cells \cite{green_enhanced_2011,yu_fundamental_2010-1}. In this regime, where the thickness of the solar cell is less than the optical wavelength, traditional ray optics does not hold, and the fundamental unanswered questions are: what is the upper bound on absorption enhancement, and what surface texture realizes this limit?
	
	In the sub-wavelength regime, there are discrete propagating modes (i.e. modes that are totally internally reflected), which can no longer be modeled as a continuum density of states. Stuart and Hall \cite{stuart_thermodynamic_1997} attempted to establish the absorption enhancement limit in the sub-wavelength by accounting for these discrete propagating modes, but they make the assumption that the introduced texture does not change the modal structure from that of a flat slab. This assumption does not hold, especially for thin solar cells where the amplitude of the texture is on the order of the thickness. In order to calculate a true limit in the sub-wavelength, the full modal structure needs to be taken into account, self-consistently. Yu et al. \cite{yu_fundamental_2010-1,yu_angular_2011,yu_thermodynamic_2012} also attempt to establish a fundamental limit in the sub-wavelength regime, but their approach depends on knowledge of the modal structure. In this work, we make no assumptions about the modal structure. We numerically find the optimal sub-wavelength surface texture by using computational inverse electromagnetic design.
	
	Our work differs from prior efforts to find the optimal surface texture for thin absorber layers in the following ways:
\begin{enumerate}[label={(\arabic*)}]
\item	Our absorber thickness is sub-wavelength, i.e. the wavelength of the light in the material is greater than the average thickness of the material. Many papers look at texturing for absorber thicknesses in the micron range \cite{yu_fundamental_2010,zeng_efficiency_2006,garnett_light_2010,shir_performance_2010,bermel_improving_2007,heine_submicrometer_1995,senoussaoui_thin-film_2004,kelzenberg_enhanced_2010,wang_absorption_2012,madzharov_influence_2011,kowalczewski_engineering_2012,eyderman_solar_2013}, a regime generally governed by ray optics.

\item	To evaluate the light trapping performance of a texture for a flat-plate, non-concentrating, non-tracking solar cell, we report the absorption enhancement averaged over frequency and over all angles in the hemisphere. A valid comparison against the ray optics limit must be angle- and frequency-averaged, instead of over a limited angular range \cite{yu_fundamental_2010,zeng_efficiency_2006,garnett_light_2010,eyderman_solar_2013,tobias_light_2008,wang_highly_2013,martins_engineering_2012,abass_dual-interface_2012,xia_misaligned_2013} or a narrowband of frequencies \cite{sheng_wavelength-selective_1983,duhring_optimization_2013}.

\item	 To derive general principles, we treat a weakly absorbing material with broadband, single-pass absorption of 1.6\%. This weak single pass absorption reveals the full benefit of light trapping. Stronger absorbance would saturate the maximum absorption enhancement possible, as seen in Refs.~\cite{senoussaoui_thin-film_2004,zhu_nanodome_2010,saeta_how_2009,zhou_photonic_2008,park_absorption_2009,eisele_periodic_2001,ferry_light_2010,fahr_approaching_2011,grandidier_light_2011,pratesi_disordered_2013}.

\item	We utilize a high index absorber material with $n = 3.5$. Though Refs.~\cite{green_enhanced_2011,yu_fundamental_2010,callahan_solar_2012} exceed the ray optics limit for sub-wavelength absorber layers, they do so for a low-index absorber ($n < 2$) sandwiched by a higher index cladding.

\item	In our optimization, we look for the most general optimal 3D texture, rather than optimizing a 2D texture (with no variation along the third dimension) \cite{duhring_optimization_2013,sheng_optimization-based_2011} or making constraints on the shape, such as optimizing 1D or 2D grating parameters \cite{yu_fundamental_2010,senoussaoui_thin-film_2004,abass_dual-interface_2012,eisele_periodic_2001} or the arrangement of nanowires \cite{kelzenberg_enhanced_2010}.

\end{enumerate}

\section*{Optimization Algorithm}

\begin{figure}[h]
\begin{center}
\includegraphics[scale=0.5]{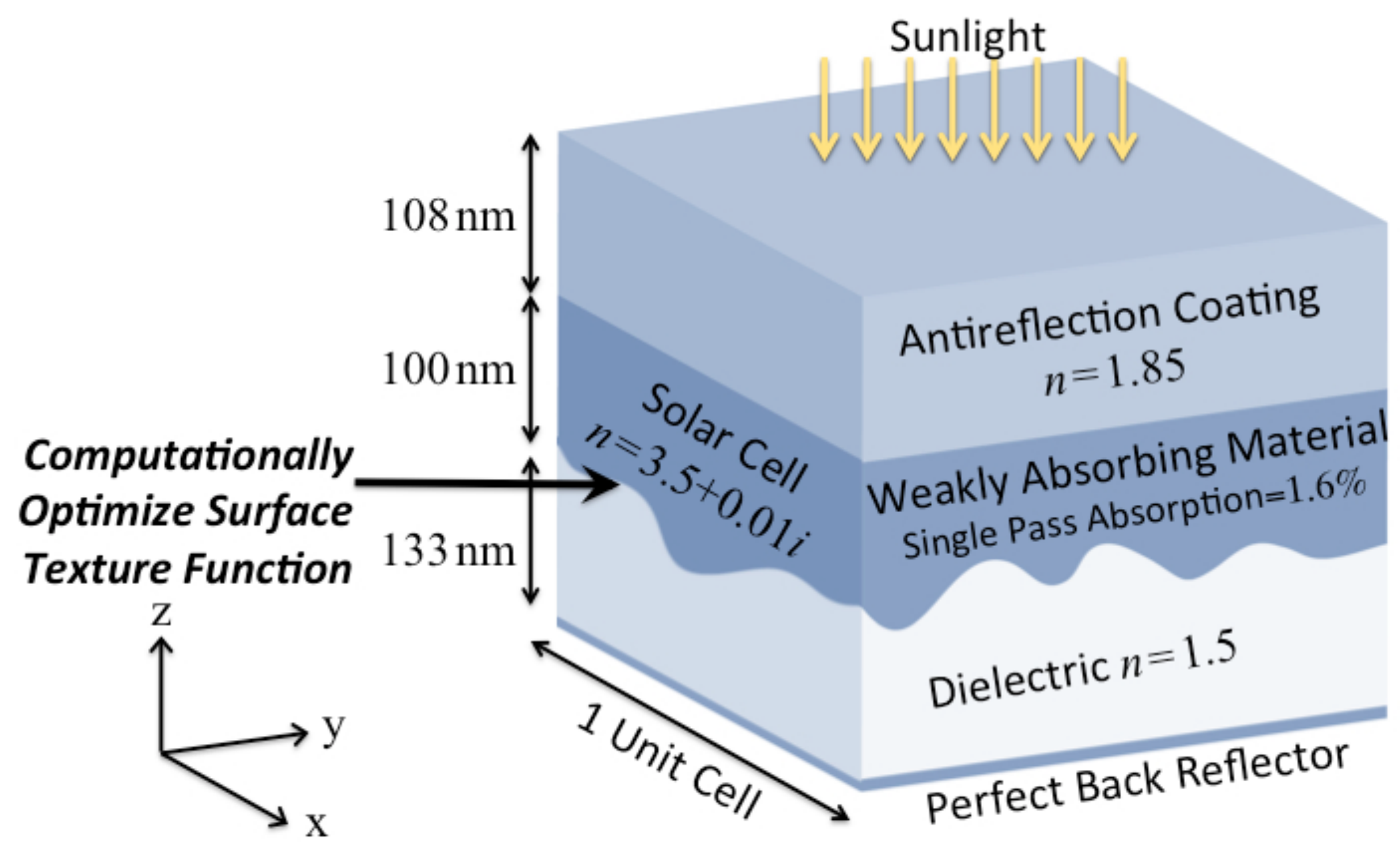}
\caption{The bottom surface texture of the absorbing material is computationally optimized. This diagram is a schematic of 1 unit cell; there are periodic boundary conditions along the $yz$ and $xz$ planes.}
\label{1setup}
\end{center}
\end{figure}

	The optimization geometry is shown in Fig.~\ref{1setup}, and is meant to be consistent with the practical requirements of a thin film solar cell. The setup consists of a weakly absorbing semiconductor material of index $n = 3.5$, with average thickness of 100 nm, and a flat top surface compatible with a conventional anti-reflection coating. The unknown texture on the bottom surface is specified within 2D periodic boundary conditions. The absorption is evaluated in the important solar frequency range, 350 THz to 400 THz (1.45 eV to 1.65 eV, or 750 nm to 860 nm free space wavelength), a bandwidth relative to center frequency of 1/8. This is a bandedge photon energy range where even a direct bandgap semiconductor like GaAs needs some absorption assistance.  We do not consider the full solar spectral bandwidth when designing a surface texture, since at most higher frequencies the direct gap absorption is sufficient. Note that Maxwell's Equations are scale-invariant, meaning that solutions described here can be scaled to different bandgaps.
	
	The average thickness of 100 nm is less than a half wavelength in the material, placing us in the subwavelength regime. An artificial weakly absorbing material ($n_{real} = 3.5$ and $\alpha=1.6 \times 10^3 \, cm^{-1}$) is chosen in order to arrive at general conclusions related to weak optical absorption. The semiconductor is specified to have a uniform $\alpha d = 0.016$ single-pass absorption throughout the band, small enough to benefit from light trapping, but large enough to allow faster numerical convergence and accuracy. A more highly absorbing material might saturate at 100\% absorption, obscuring the benefit of the surface texturing.
	
	An antireflection (AR) coating is applied to the top of the solar cell structure: it is fixed at a quarter wavelength (108 nm) for the center wavelength in the optimization bandwidth, with $n_{AR} = \sqrt{n_{air} \times n_{absorber}} = 1.85$. A bottom surface texture was chosen for the absorber layer so we can keep the antireflection coating fixed in our optimization algorithm. Beneath the absorber layer is a non-absorbing back dielectric layer of $n = 1.5$ (adjusted to 133 nm average thickness) followed by a perfect back reflector.
	
	The periodic surface texture function, $h$, is represented by a truncated Fourier series: 
\begin{equation}\label{eq:fourier} h(x,y) = \sum _{m=-2}^2 \sum _{n=-2}^2 c_{mn}e^{\frac{m 2 \pi x}{\Lambda }}e^{\frac{n 2 \pi y}{\Lambda }},\end{equation}
where $\Lambda$ is the periodicity, and $c_{mn}$ are the Fourier coefficients. In our optimization algorithm, we keep the periodicity and the zeroth order Fourier coefficient (the average absorber layer thickness) fixed, and allow the other Fourier coefficients to evolve. We truncate the Fourier series to avoid small highly resonant features that would not be robust in manufacturing. 
	
	Our optimization algorithm scripts are written in MATLAB, following the procedure described in Ref.~\cite{miller_photonic_2012}. Our optimization uses an adjoint gradient method to search for a local optimum \cite{giles_introduction_2000}. To find the absorption of the solar cell, we simulate the solar cell structure of Fig.~\ref{1setup} in ``Lumerical FDTD Solutions," a commercial finite-difference time-domain solver for Maxwell's Equations, evaluating the absorption at 30 points within the frequency bandwidth. Each iteration takes approximately 15 minutes on our computational cluster of 128 cores, and the optimization converges after about 25 iterations.
	
	The selection of the Figure of Merit is critical. We maximize the absorption enhancement at the frequency with the lowest absorption, a minimax Figure of Merit \cite{murray_projected_1980}, which allows us to achieve good absorption over the whole frequency band. In one iteration, we evaluate the absorption for each frequency, at each of the two perpendicular polarizations of normally incident light. We then take the lowest absorption as the Figure of Merit. At the end of the optimization, we compute the angle-averaged performance. The Lambertian angle-averaged performance as a function of frequency is given by:
\begin{equation}\label{eq:angleAve} \int _0^{2\pi }\int _0^{\frac{\pi }{2}} AE(\theta ,\phi )\times \cos (\theta ) \times \sin (\theta ) d\theta d\phi ,\end{equation}
where $AE$ is the absorption enhancement found by dividing the absorption by the average single pass absorption $\alpha d = 0.016$, and averaging over the two perpendicular polarizations. At the end of the full optimization, we evaluate Eq.~\ref{eq:angleAve} by simulating 12 angles over the hemisphere, with two orthogonal polarizations for every angle. 

\section*{Results and Discussion}

\begin{figure}[htbp]
\begin{center}
\includegraphics[scale=0.5]{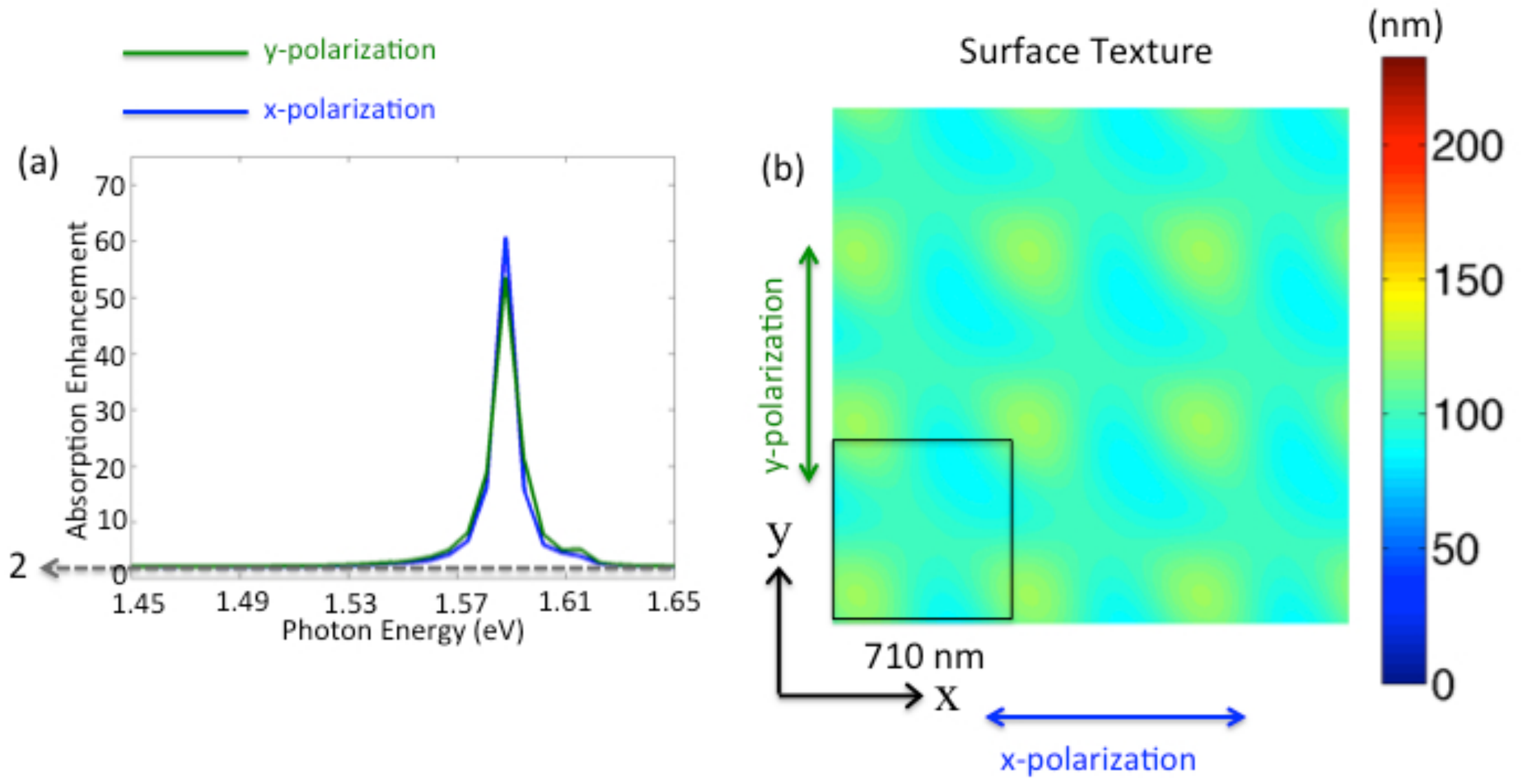}
\caption{(a) The initial absorption enhancement as a function of frequency and (b) a top-down view of the surface texture; the colors show the height of the absorbing material (from the antireflection coating to the bottom dielectric, as seen in Fig.~\ref{1setup}).}
\label{2initialSample}
%\end{center}
%\end{figure}

%\begin{figure}[h]
%\begin{center}
\includegraphics[scale=0.5]{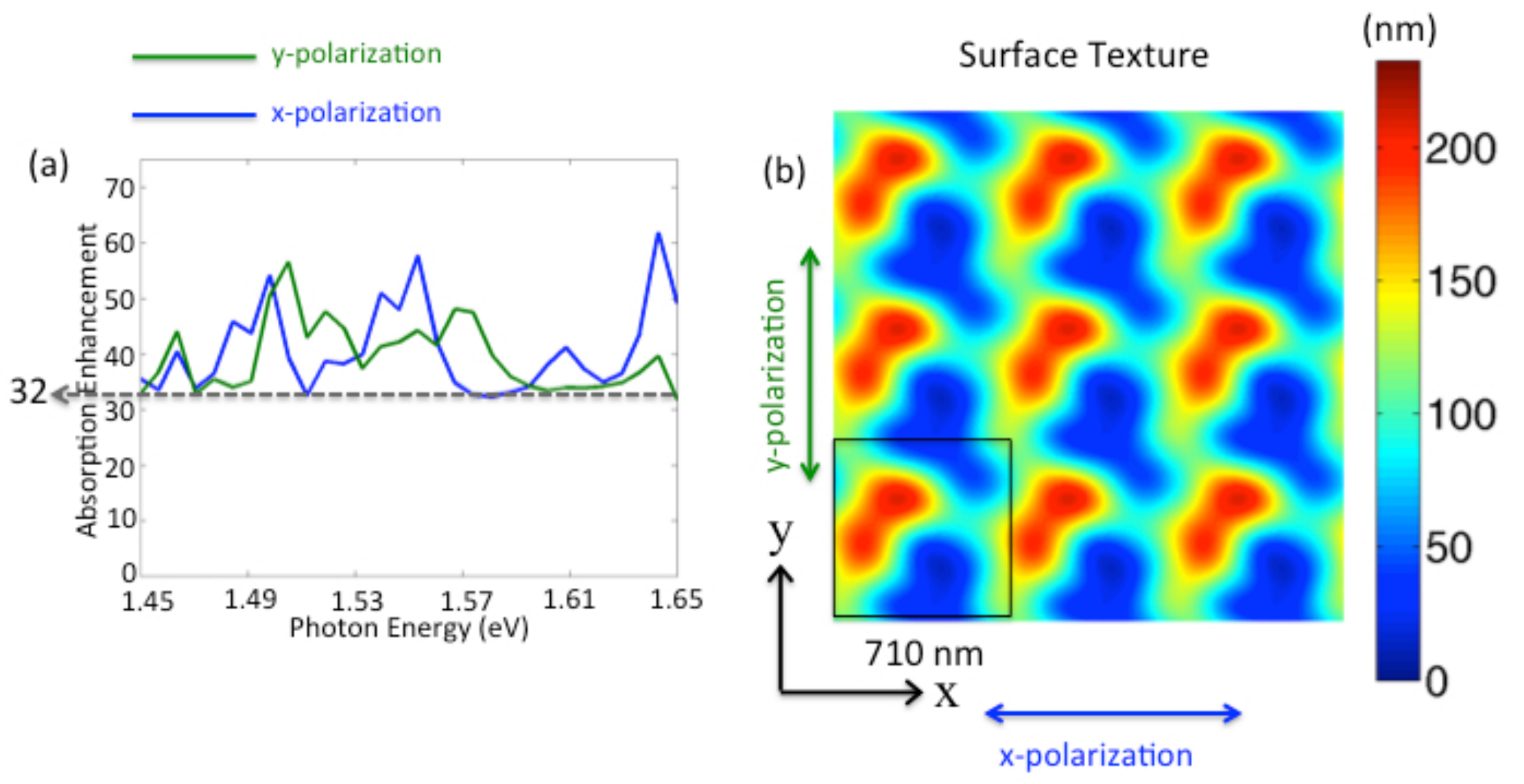}
\caption{(a) The final absorption enhancement as a function of frequency and (b) a top-down view of the surface texture; the colors show the height of the absorbing material (from the antireflection coating to the bottom dielectric, as seen in Fig.~\ref{1setup}).}
\label{3finalSample}
\end{center}
\end{figure}

\begin{figure}[htbp]
\begin{center}
\includegraphics[scale=0.5]{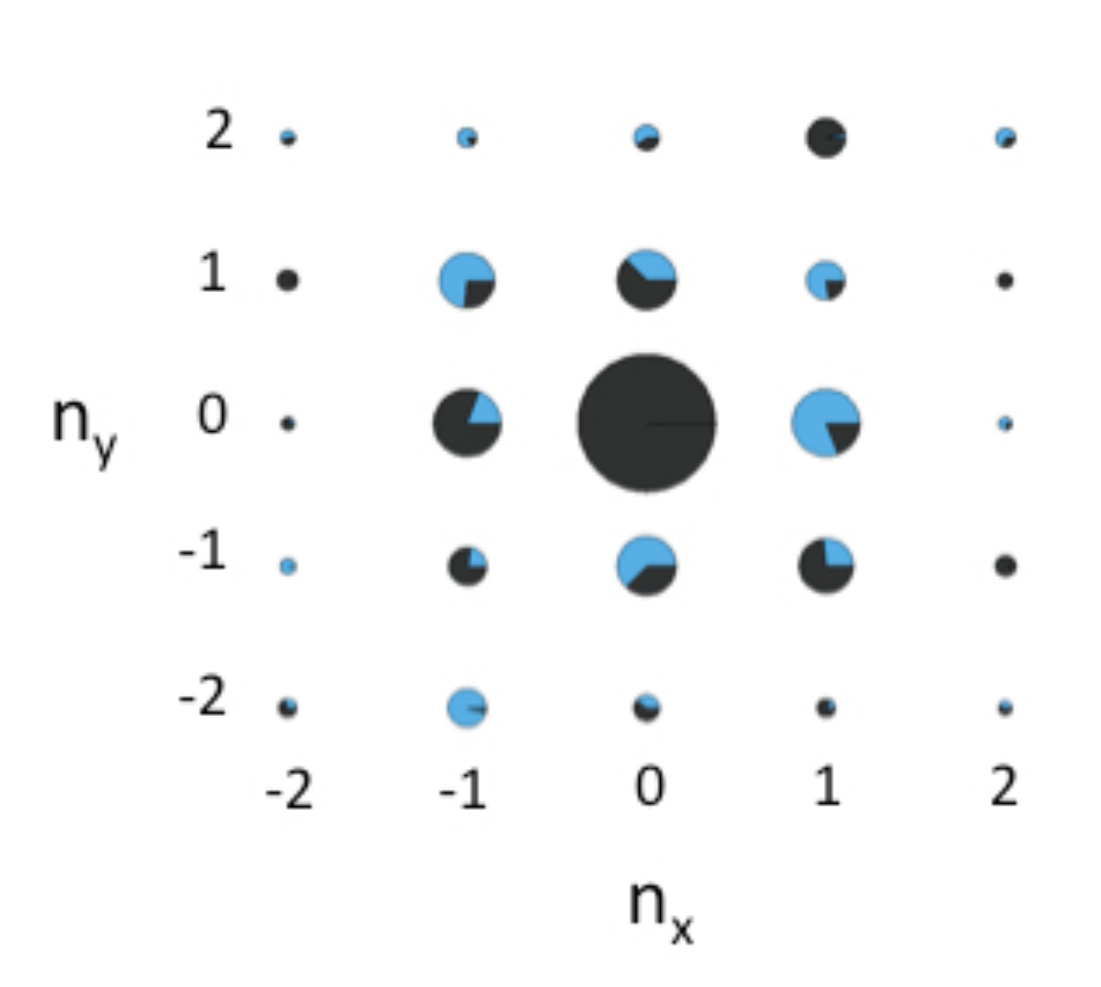}
\caption{The reciprocal (k-) space representation for the final texture seen in Fig.~\ref{3finalSample}. The blue pie slices represent the phase of the complex exponential Fourier coefficients. }
\label{4kspaceSample}
\end{center}
\end{figure}

\begin{figure}[htbp]
\begin{center}
\includegraphics[scale=0.4]{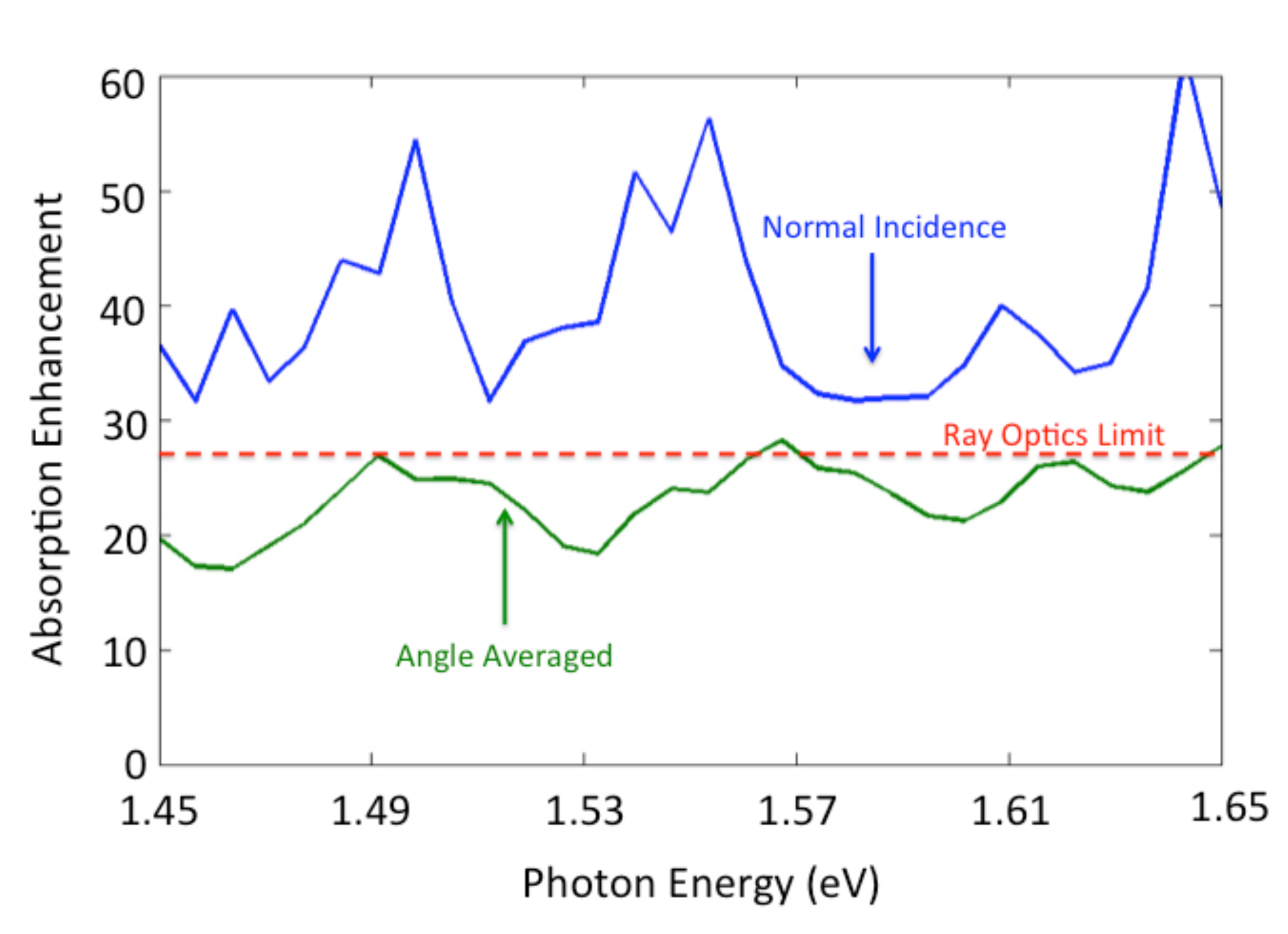}
\caption{The absorption enhancement for the texture in Fig.~\ref{3finalSample}, plotted as a function of frequency at normal incidence (blue) and angle averaged (green). }
\label{5FOMSample}
\end{center}
\end{figure}

	We started the algorithm from noisy initial conditions with fixed periodicity of 710 nm ($=  3.1\lambda_n = 3.5$). We randomly picked initial Fourier coefficients in the range of 0 to 8 nm. In this first example, we achieved a minimum absorption enhancement $AE = 32$ for a 100 nm average thickness absorber layer at normal incidence. The progression of the surface texture and absorption enhancement at normal incidence from the first iteration to the last is shown in Figs.~\ref{2initialSample} and \ref{3finalSample}. The reciprocal space representation (the magnitudes and phases of the Fourier coefficients) of the final surface is shown in Fig.~\ref{4kspaceSample}. The effect of our minimax Figure of Merit in optimizing for the lowest absorbing frequency and for achieving high absorption over the full band can be seen in this progression. Resonant peaks from the initial case flatten out, and both the minimum and average absorption enhancement improve. The angle-averaged performance is shown in Fig.~\ref{5FOMSample}. Angle- and frequency-averaged, this texture achieves an absorption enhancement of $AE = 23$ relative to 1.6\% single pass absorption.
	
\begin{figure}[htbp]
\begin{center}
\includegraphics[scale=0.5]{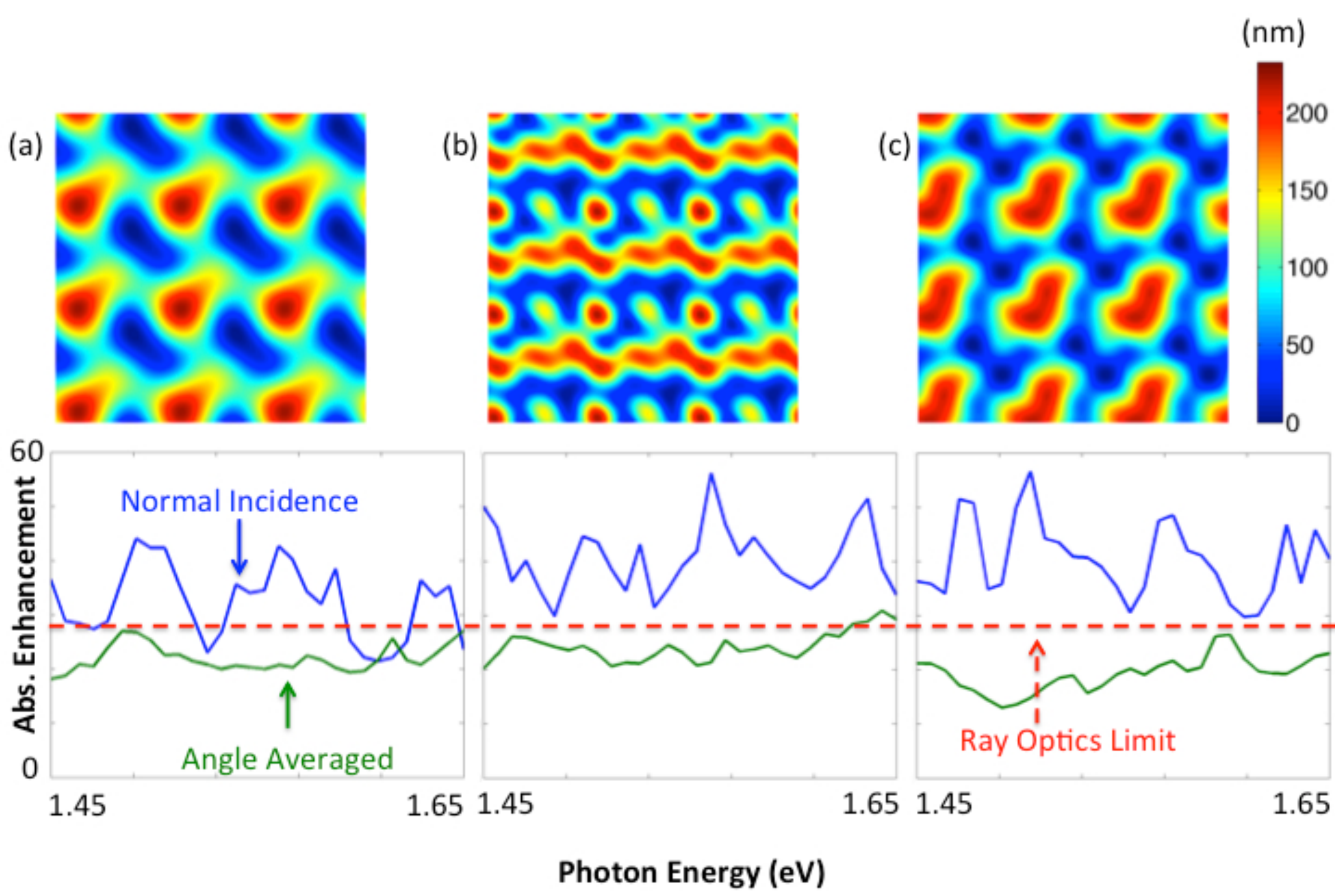}
\caption{The surface textures and absorption enhancement as a function of frequency for different initial conditions, revealing a broad optimum.}
\label{6broadOpt}
\end{center}
\end{figure}

\begin{figure}[htbp]
\begin{center}
\includegraphics[scale=0.5]{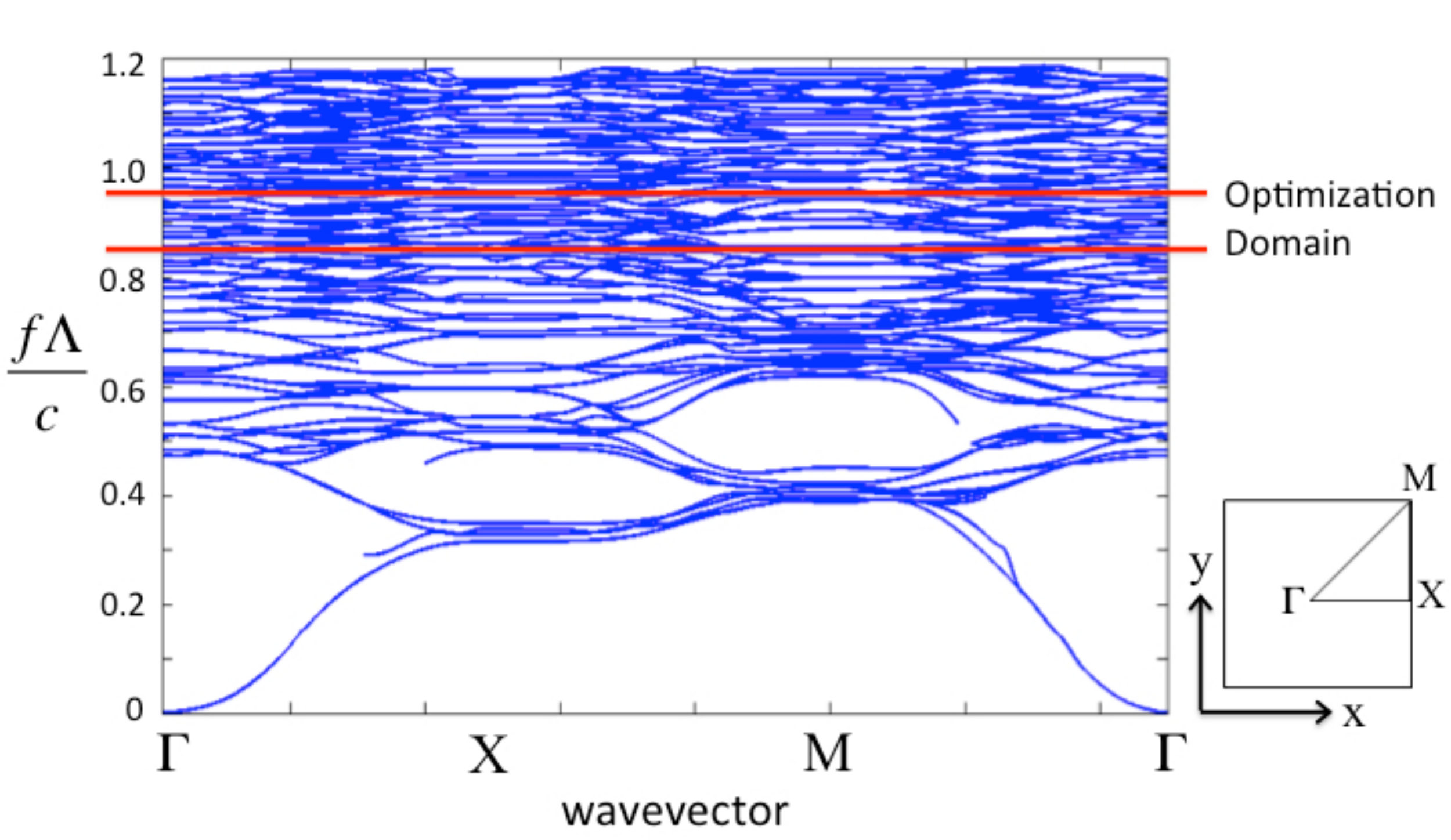}
\caption{The photonic bandstructure for the texture in Fig.~\ref{6broadOpt}(b).}
\label{7bandstructure}
\end{center}
\end{figure}

	Our optimization algorithm is sensitive to initial conditions; Fig.~\ref{6broadOpt} shows three cases of the final texture and final absorption enhancement both at normal incidence and angle averaged for different initial conditions. From different initial conditions, we obtain different textures reaching similar angle- and frequency-averaged absorption enhancements of $AE = $ 22, 24, and 19. The best angle- and frequency-averaged absorption enhancement of $AE = 24$ is seen in the texture in Fig.~\ref{6broadOpt}(b). A common feature of the textures is large height amplitude. The full amplitudes for the textures in Figs. \ref{3finalSample}, \ref{6broadOpt}(a), \ref{6broadOpt}(b), and \ref{6broadOpt}(c) are $\Delta h =$ 196 nm, 224 nm, 218 nm, and 217 nm, respectively. The photonic bandstructure of the optimized texture from Fig.~\ref{6broadOpt}(b) is shown in Fig.~\ref{7bandstructure}. The optimization domain (the bandwidth of frequencies that we simulate) is highlighted in Fig.~\ref{7bandstructure}. The bandstructure visually shows the need for a high modal density in the optimization domain; the optimization needs modes for the incident light to couple to.
	
\begin{figure}[htbp]
\begin{center}
\includegraphics[scale=0.5]{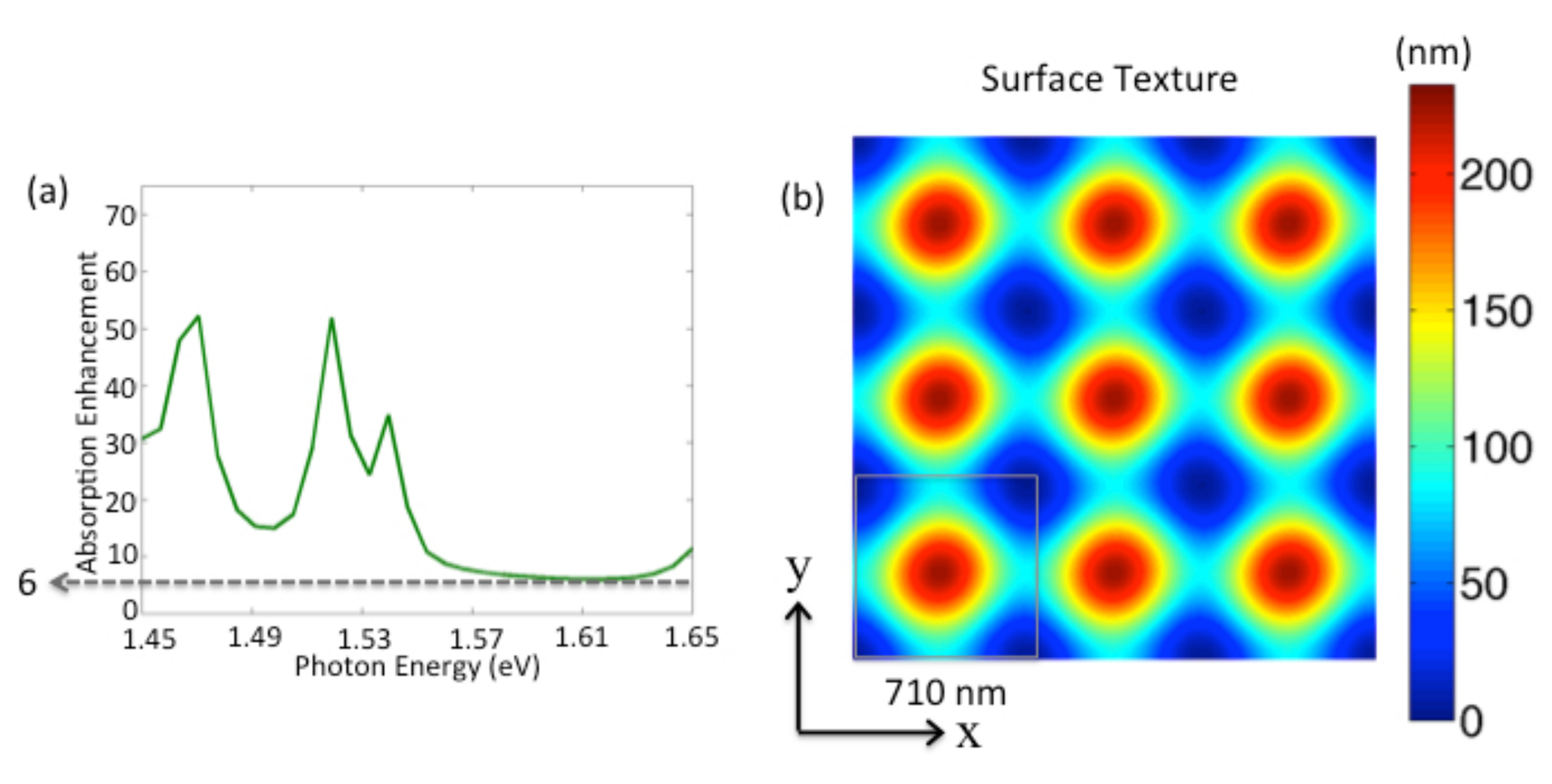}
\caption{(a) The initial absorption enhancement as a function of frequency and (b) a top-down view of the surface texture, for a symmetric texture with a slight perturbation along the diagonal. }
\label{8SymBreakInit}
%\end{center}
%\end{figure}

%\begin{figure}[htbp]
%\begin{center}
\includegraphics[scale=0.5]{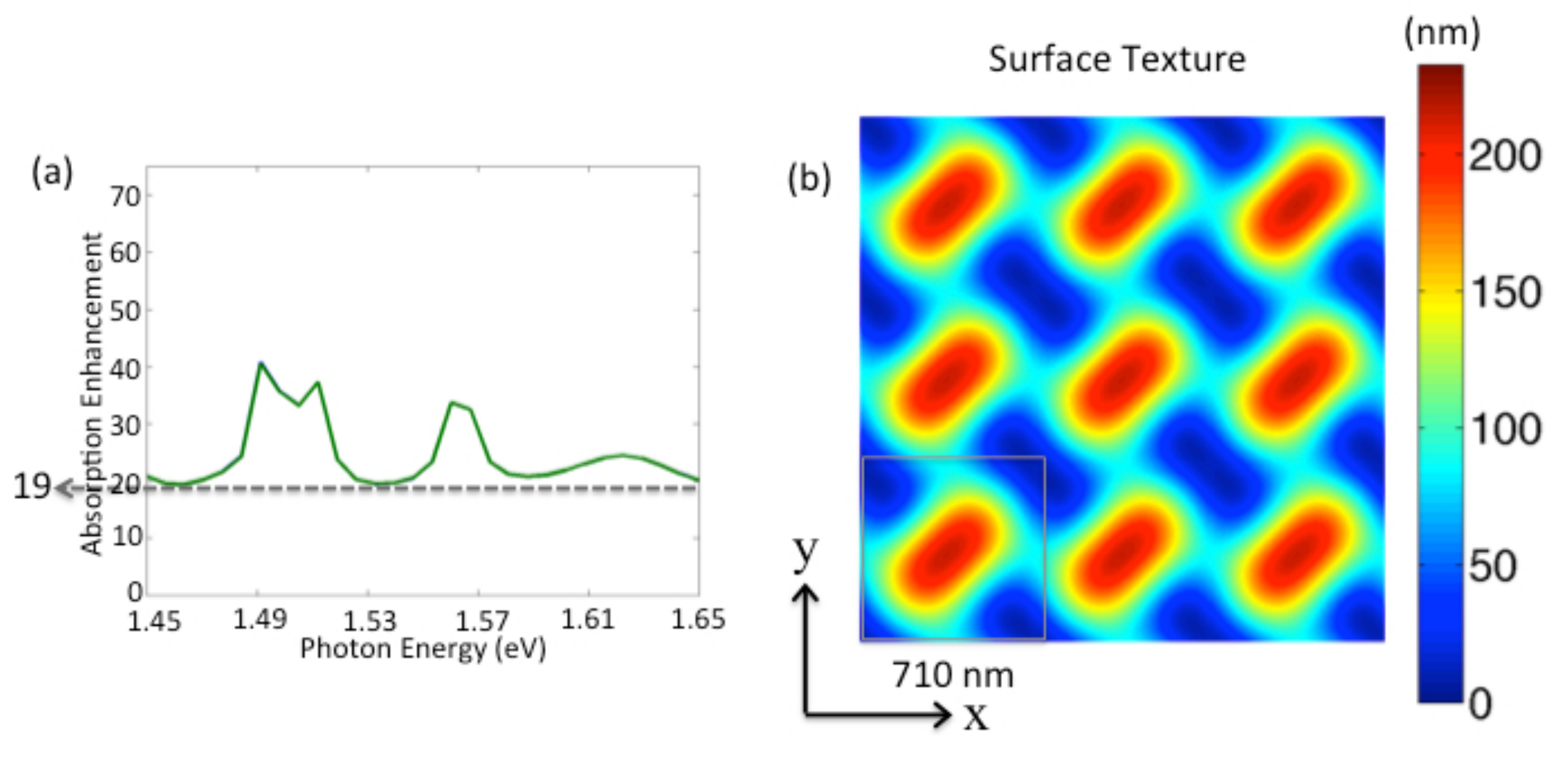}
\caption{(a) The final absorption enhancement as a function of frequency and (b) a top-down view of the surface texture, showing broken mirror symmetry, from almost symmetric initial conditions seen in Fig.~\ref{8SymBreakInit}.}
\label{9SymBreakFinal}
\end{center}
\end{figure}

\begin{figure}[htbp]
\begin{center}
\includegraphics[scale=0.5]{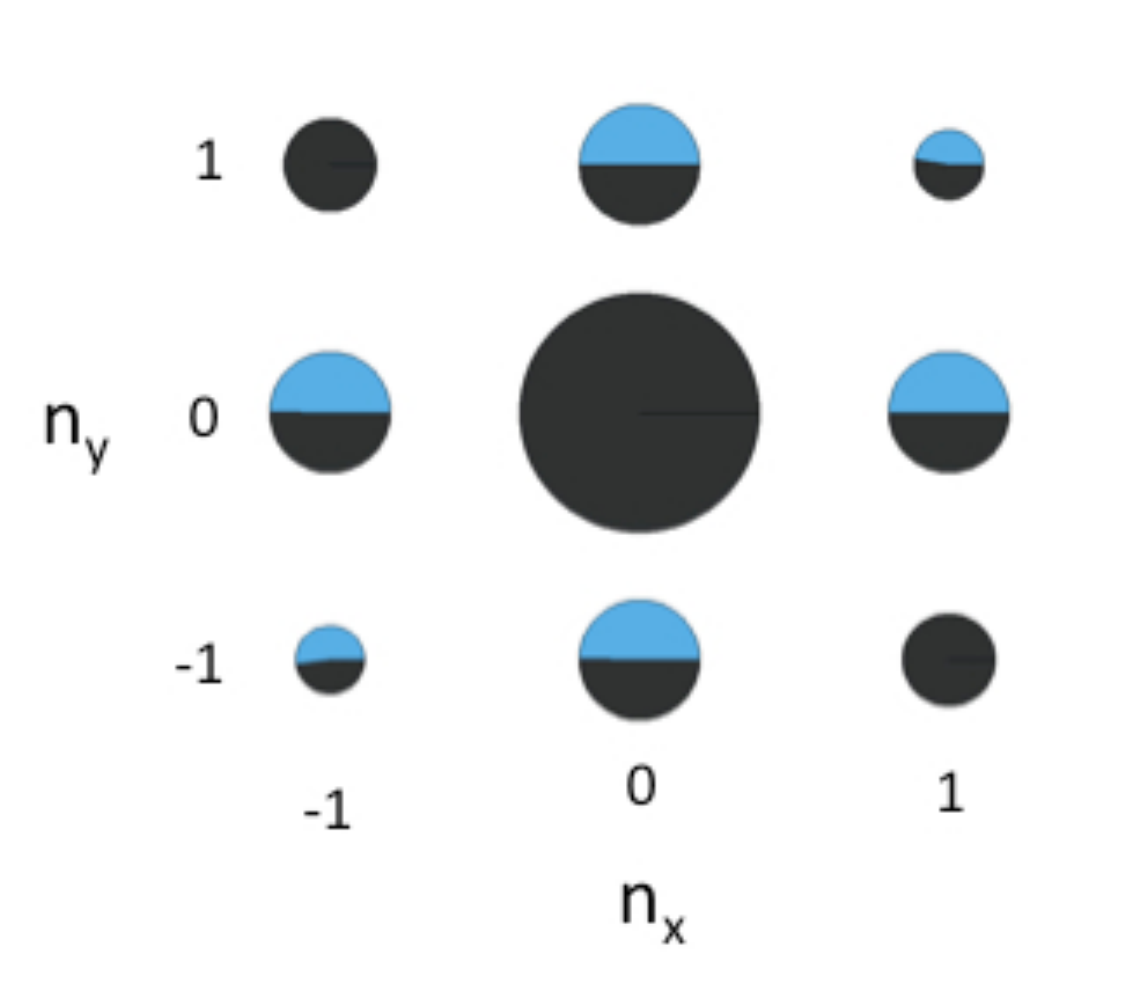}
\caption{The reciprocal space representation for the texture with broken mirror symmetry in Fig.~\ref{9SymBreakFinal}. The blue pie slices represent the phase of the complex exponential Fourier coefficients. }
\label{10SymBreakKSpace}
\end{center}
\end{figure}

	The second common feature we observe is asymmetry within the unit cell. The optimal structures appear to break the inherent mirror symmetries of the problem, with a feature growing along one of the diagonals. To demonstrate that this symmetry breaking is not an artifact of the starting noise, we started another optimization from initial symmetrized conditions, with a slight perturbation along the diagonal, as shown in Fig.~\ref{8SymBreakInit}. The result of the algorithm is shown 15 iterations later in Fig.~\ref{9SymBreakFinal} (reciprocal space diagram in Fig.~\ref{10SymBreakKSpace}). We see that this perturbation has been amplified along the diagonal, suggesting that symmetry breaking is a fundamental feature of optimal textures. There appears to be no significance to the direction of the asymmetric component: the symmetry will break in the opposite direction (along $x = -y$) if the initial perturbation is in that direction.
	
\begin{figure}[htbp]
\begin{center}
\includegraphics[scale=0.4]{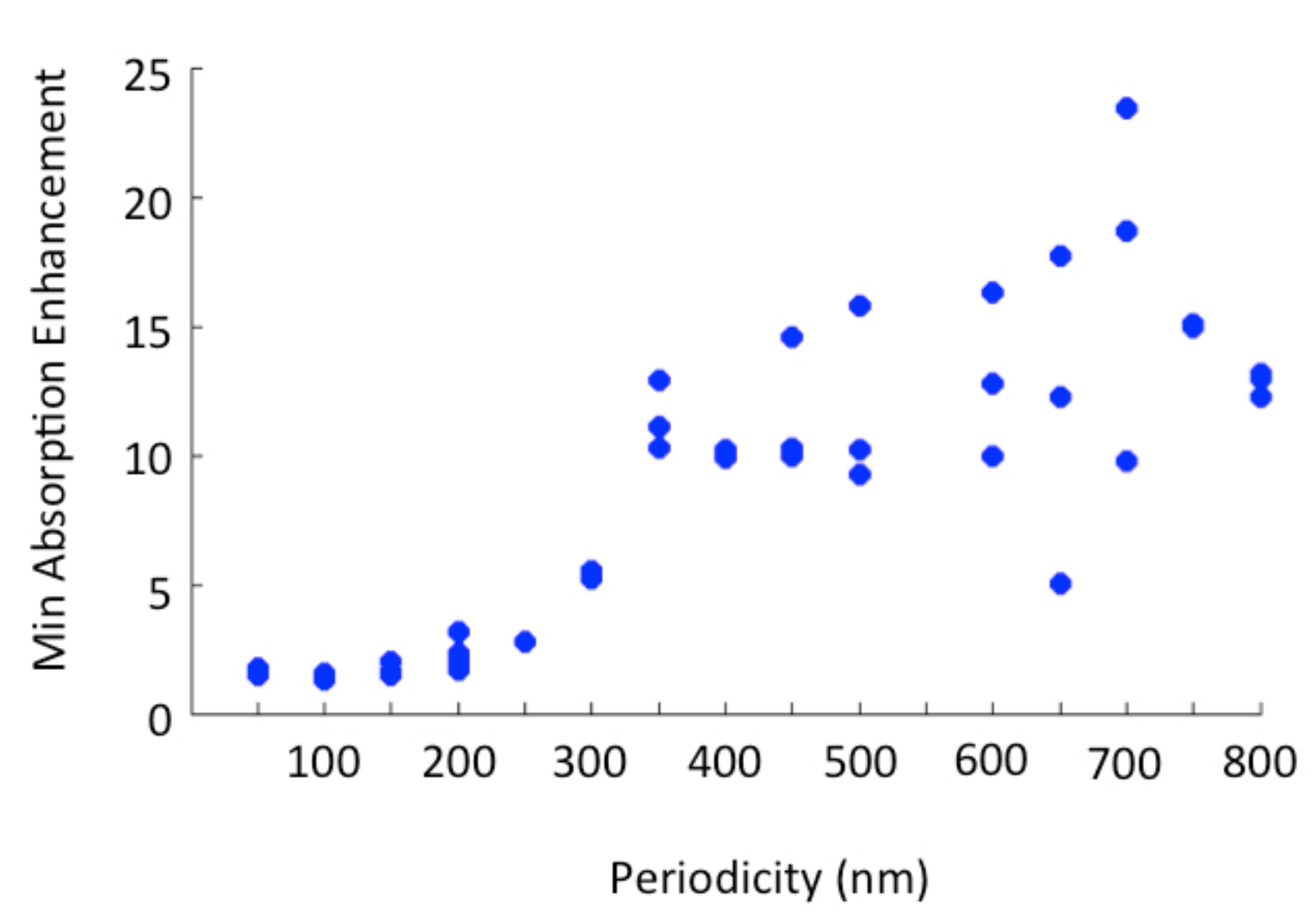}
\caption{Optimizations were carried out at periodicities from 50 to 800 nm, in increments of 50 nm. For each periodicity, at least 3 optimizations were completed for randomly chosen initial starting noise. The Figure of Merit (absorption enhancement for the worst performing frequency and polarization at normal incidence) is plotted for each optimization.}
\label{11PeriodSweep}
\end{center}
\end{figure}

	In our optimizations, we kept the periodicity fixed at 710 nm. To find the optimal periodicity, we ran a sweep of optimizations with fixed periodicities from 50 nm to 800 nm in increments of 50 nm. Fig.~\ref{11PeriodSweep} plots the Figure of Merit (absorption enhancement at the minimum performing frequency) achieved in these optimizations. The smaller periodicities did not optimize well; periodicities less than 350 nm did not achieve minimum absorption enhancement $AE > 10$. The best optimizations occurred at 700 nm; perhaps because this periodicity brought the optimization frequency band high into the photonic bandstructure where the optical density of states is large. 

\begin{figure}[htbp]
\begin{center}
\includegraphics[scale=0.5]{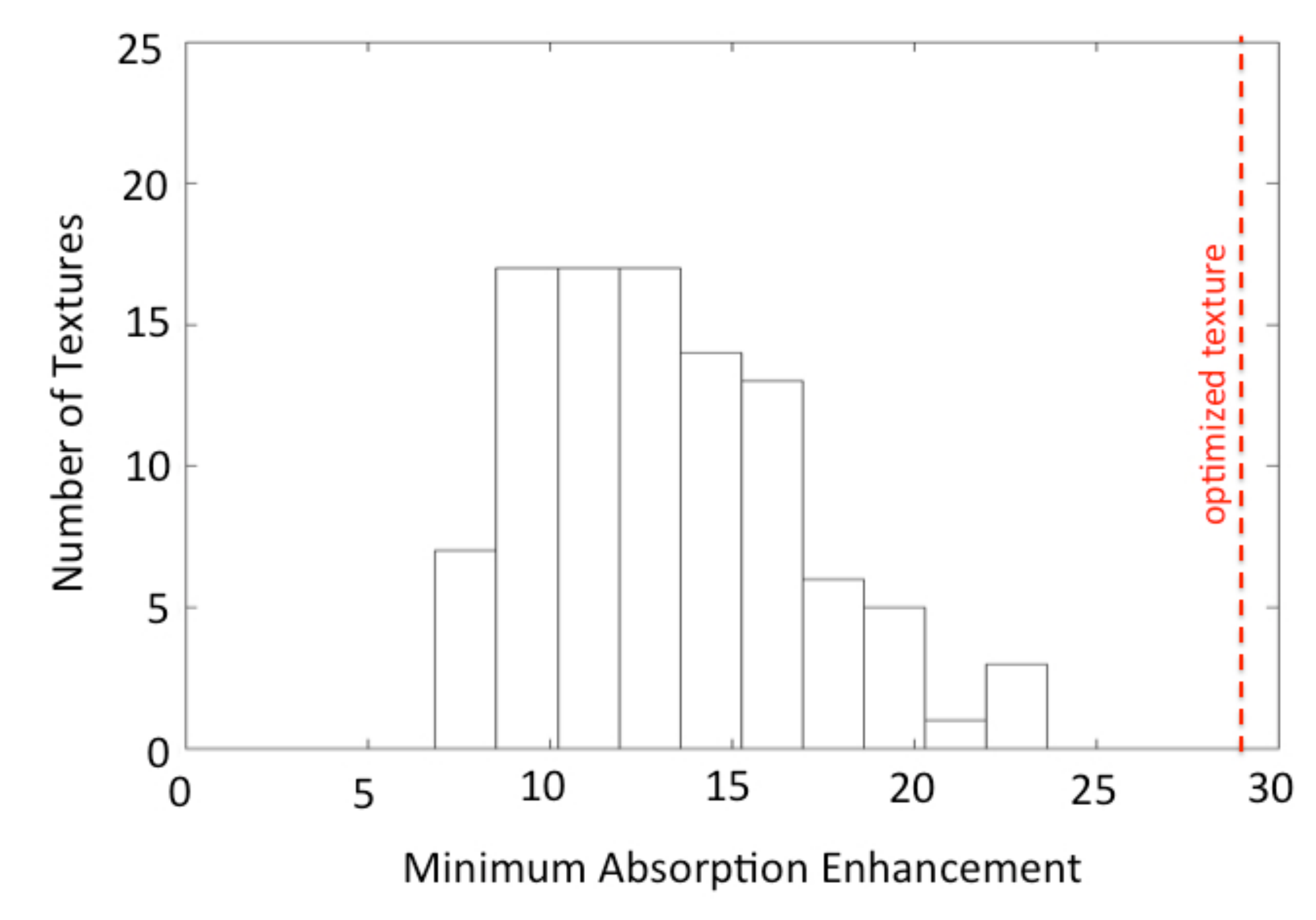}
\caption{The Figure of Merit (minimum absorption enhancement at normal incidence) plotted for 100 randomly generated textures of 710 nm periodicity. For comparison, the Figure of Merit for the optimized texture in Fig.~\ref{6broadOpt}(b) is shown by the dotted red line.}
\label{12RandomTex100}
\end{center}
\end{figure}

\begin{figure}[htbp]
\begin{center}
\includegraphics[scale=0.5]{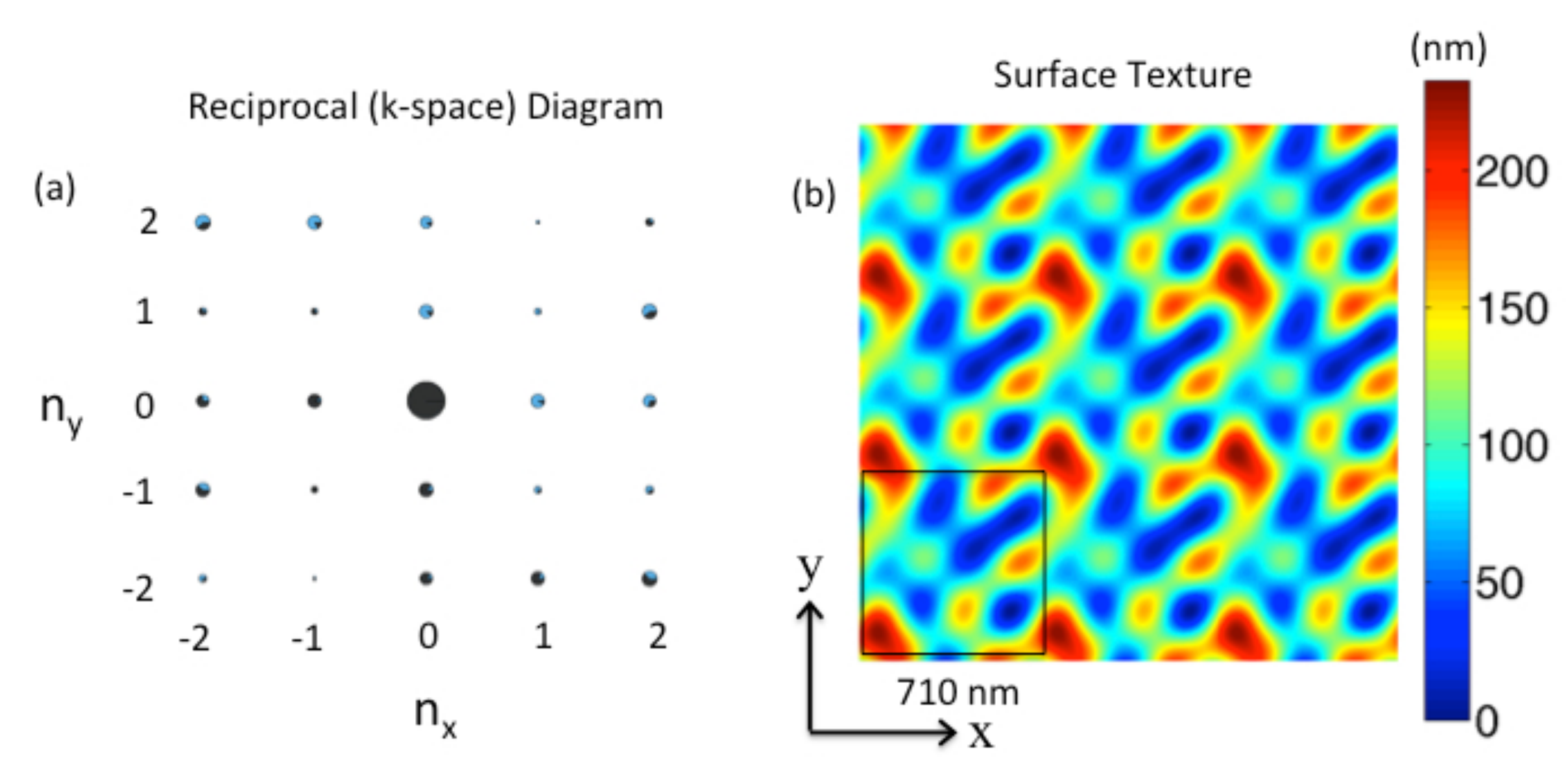}
\caption{The (a) reciprocal k-space diagram and (b) real-space top down view of the median randomly generated texture from Fig.~\ref{12RandomTex100}.}
\label{13RandomTex710Med}
\end{center}
\end{figure}

\begin{figure}[htbp]
\begin{center}
\includegraphics[scale=0.5]{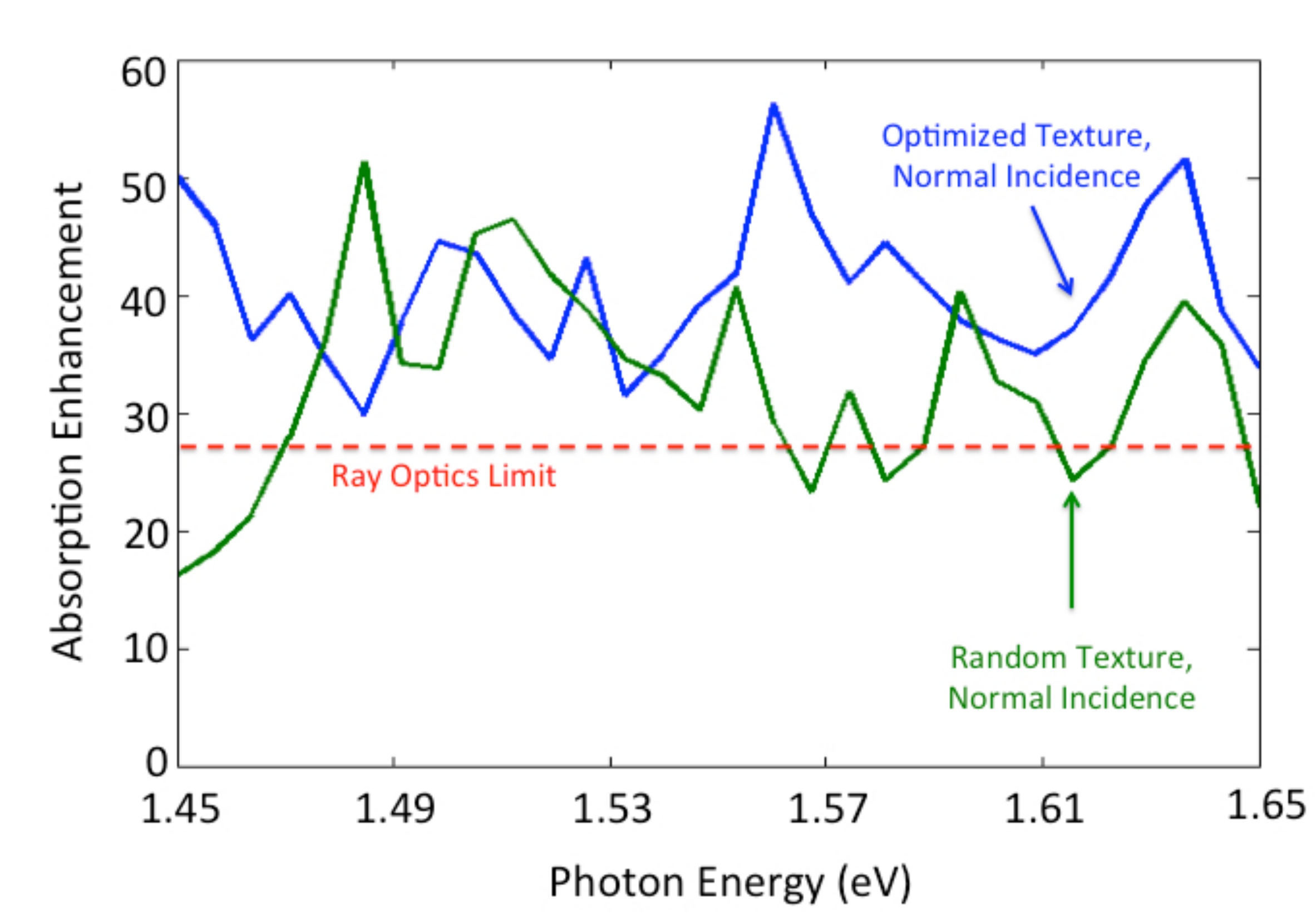}
\caption{The absorption enhancement as a function of frequency at normal incidence for the optimized texture from Fig.~\ref{6broadOpt}(b) (blue) compared with the median random texture from Fig.~\ref{13RandomTex710Med} (green). Lines are averaged over the two orthogonal polarizations.}
\label{14Rand710CompareNorm}
%\end{center}
%\end{figure}

%\begin{figure}[htbp]
%\begin{center}
\includegraphics[scale=0.5]{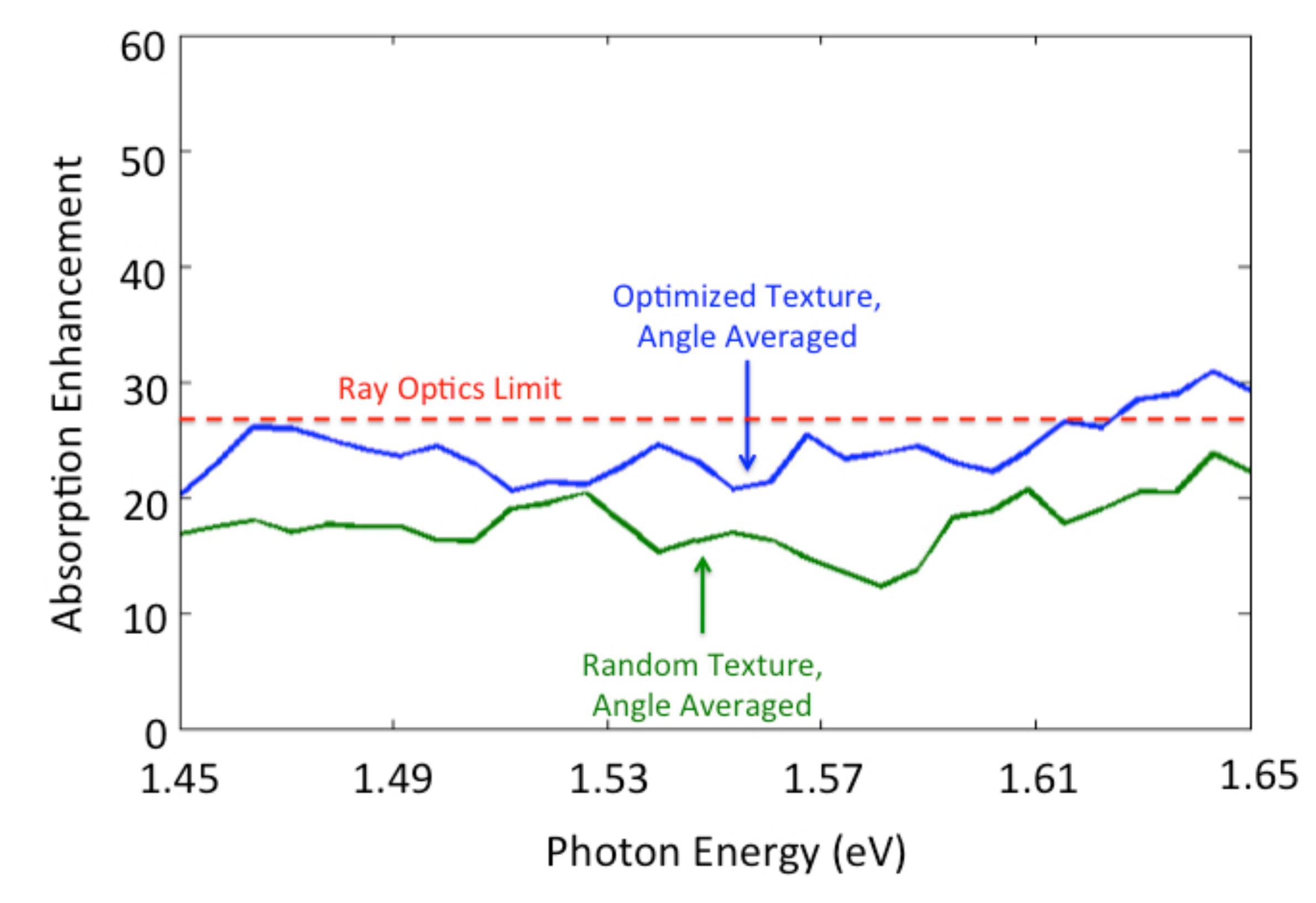}
\caption{The absorption enhancement as a function of frequency, angle averaged, for the optimized texture from Fig.~\ref{6broadOpt}(b) (blue) compared with the median random texture from Fig.~\ref{13RandomTex710Med} (green).}
\label{15Rand710CompareAngle}
\end{center}
\end{figure}

	Since the optimum is not unique, it is possible that any randomly generated pattern with large amplitude could achieve a similar Figure of Merit. To check this, we randomly generated Fourier coefficients for 100 textures with a periodicity of 710 nm, with amplitudes ranging from $\Delta h = $ 223 to 233 nm, and simulated the absorption of these structures. The Figure of Merits (the lowest absorption as a function of frequency at normal incidence, for the worst performing polarization) of these random structures are plotted in Fig.~\ref{12RandomTex100}, and the random texture with the median Figure of Merit $AE = 13$ is shown in Fig.~\ref{13RandomTex710Med}. In Figs.~\ref{14Rand710CompareNorm} and \ref{15Rand710CompareAngle}, the random texture with the median Figure of Merit is compared with the optimized texture shown in Fig.~\ref{6broadOpt}(b). The Figure of Merit for the optimized texture is over two times greater than the median Figure of Merit in the randomly generated patterns. A comparison of angle- and frequency-averaged absorption performance (shown in Fig.~\ref{15Rand710CompareAngle}) shows a 33\% increase in the optimized textureÕs absorption enhancement over the median randomly generated pattern. 

\begin{figure}[htbp]
\begin{center}
\includegraphics[scale=0.45]{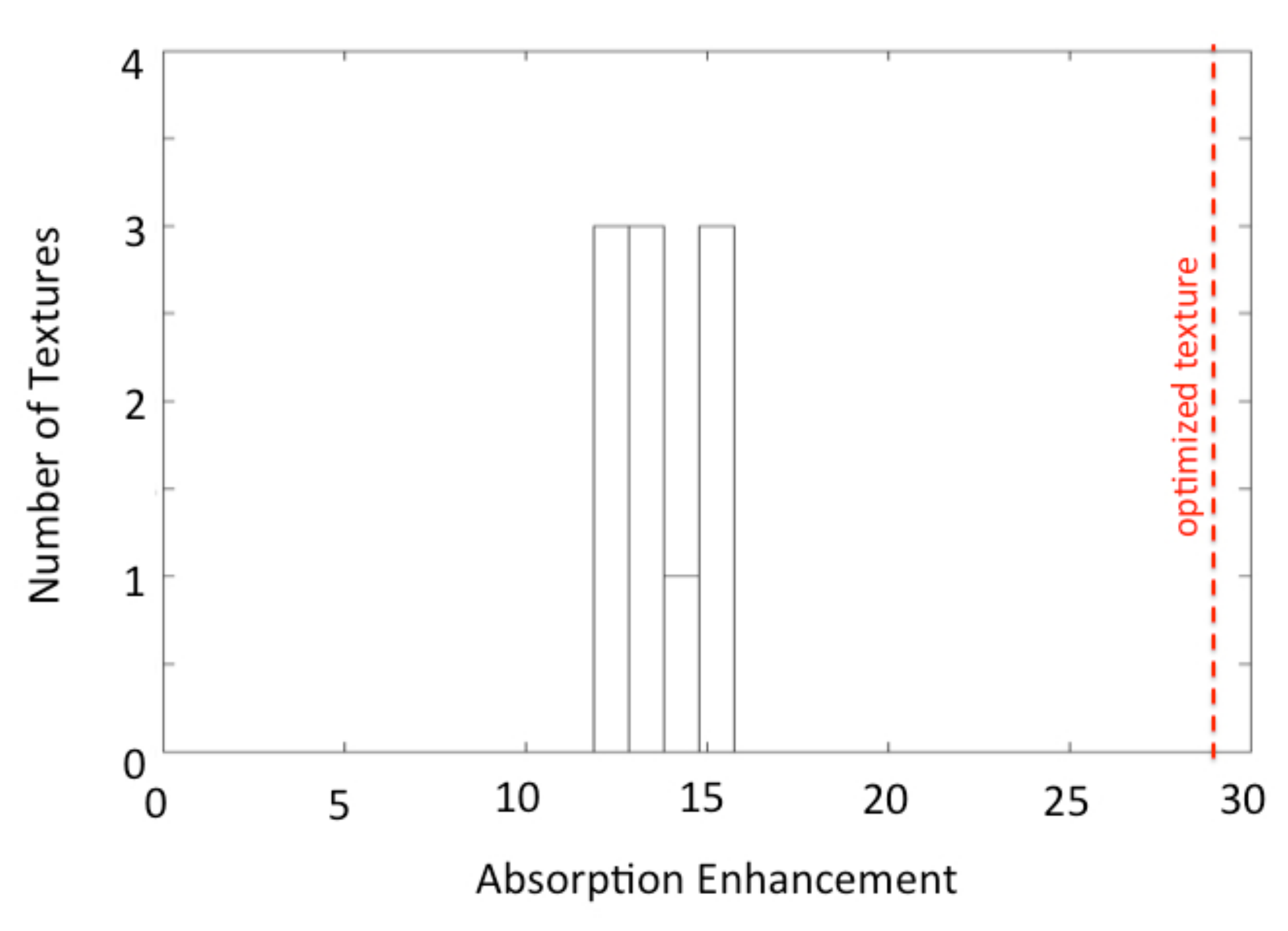}
\caption{The Figure of Merit (minimum absorption enhancement at normal incidence) plotted for 11 randomly generated textures of 2300 nm periodicity. For comparison, the Figure of Merit for the optimized texture in Fig.~\ref{6broadOpt}(b) is shown by the dotted red line.}
\label{16RandTex11}
%\end{center}
%\end{figure}

%\begin{figure}[htbp]
%\begin{center}
\includegraphics[scale=0.4]{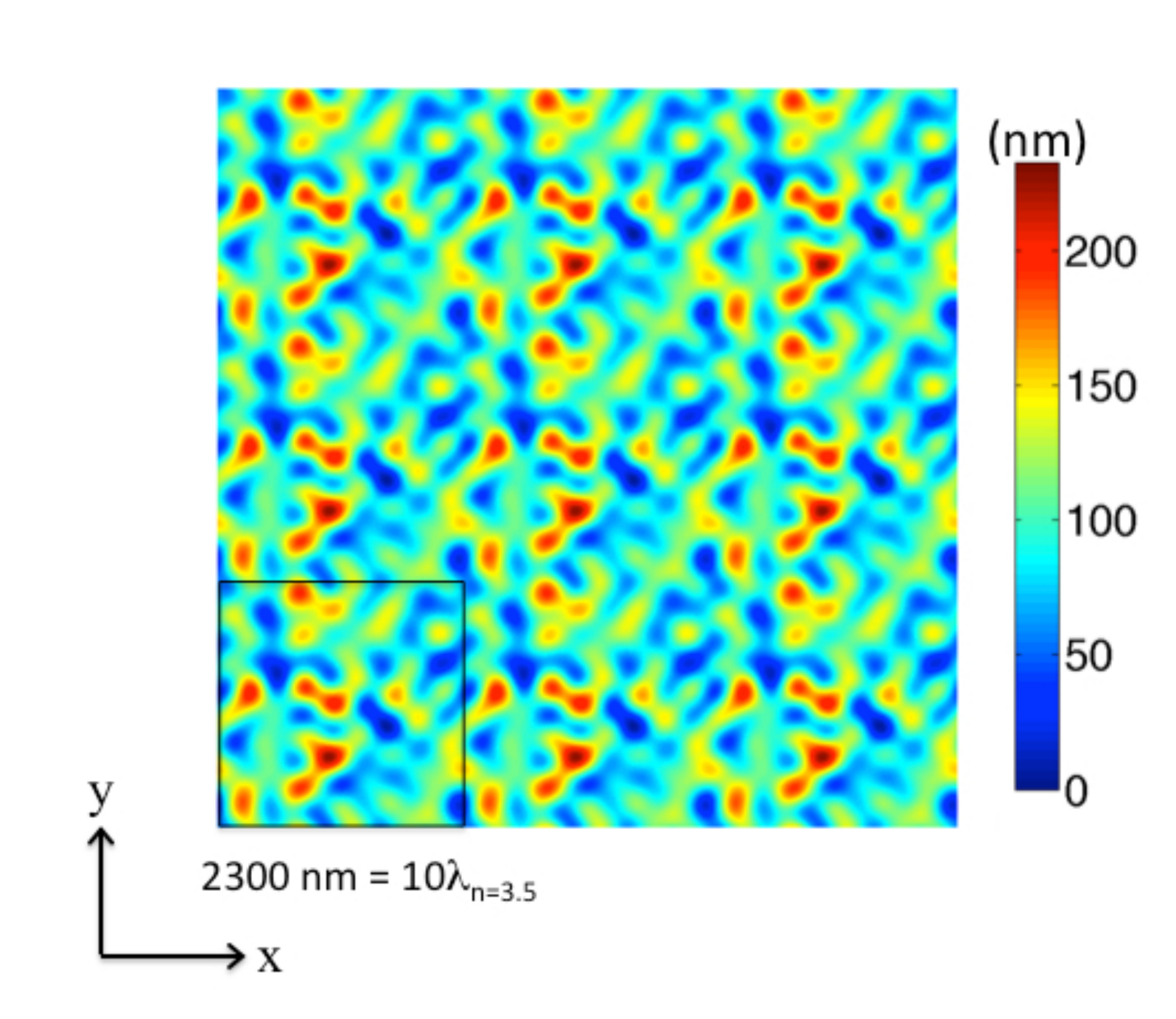}
\caption{A top-down view of the surface texture, for the randomly generated texture with periodicity of 2300 nm = $10 \lambda_n$ = 3.5, with median Figure of Merit (minimum absorption enhancement).}
\label{17RandSuperCell}
\end{center}
\end{figure}

\begin{figure}[htbp]
\begin{center}
\includegraphics[scale=0.5]{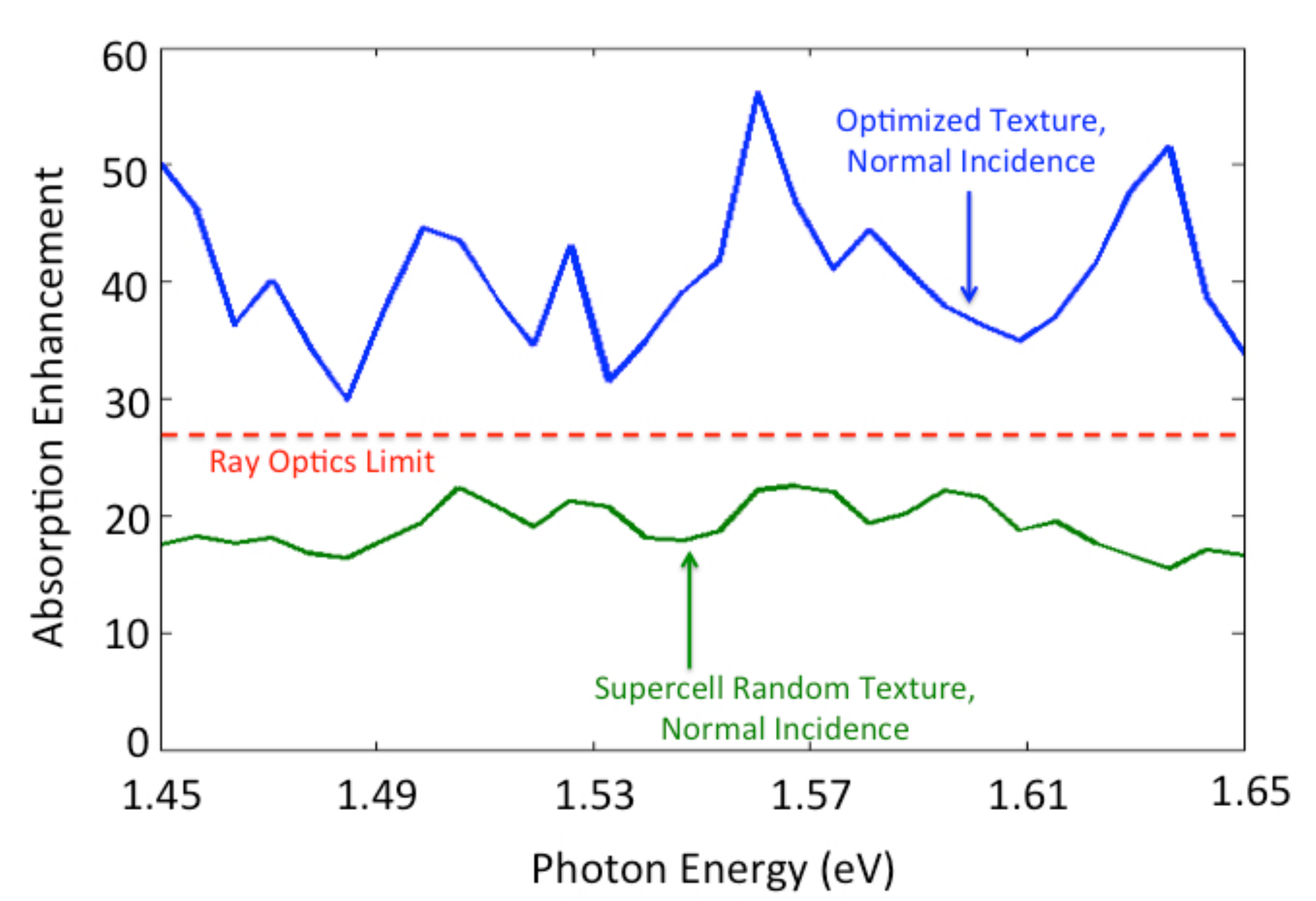}
\caption{The absorption enhancement as a function of frequency at normal incidence for the optimized texture from Fig.~\ref{6broadOpt}(b) (blue) compared with the median random texture with 2300 nm periodicity from Fig.~\ref{17RandSuperCell} (green). Lines are averaged over the two orthogonal polarizations.}
\label{18SupercellCompareNorm}
%\end{center}
%\end{figure}

%\begin{figure}[htbp]
%\begin{center}
\includegraphics[scale=0.5]{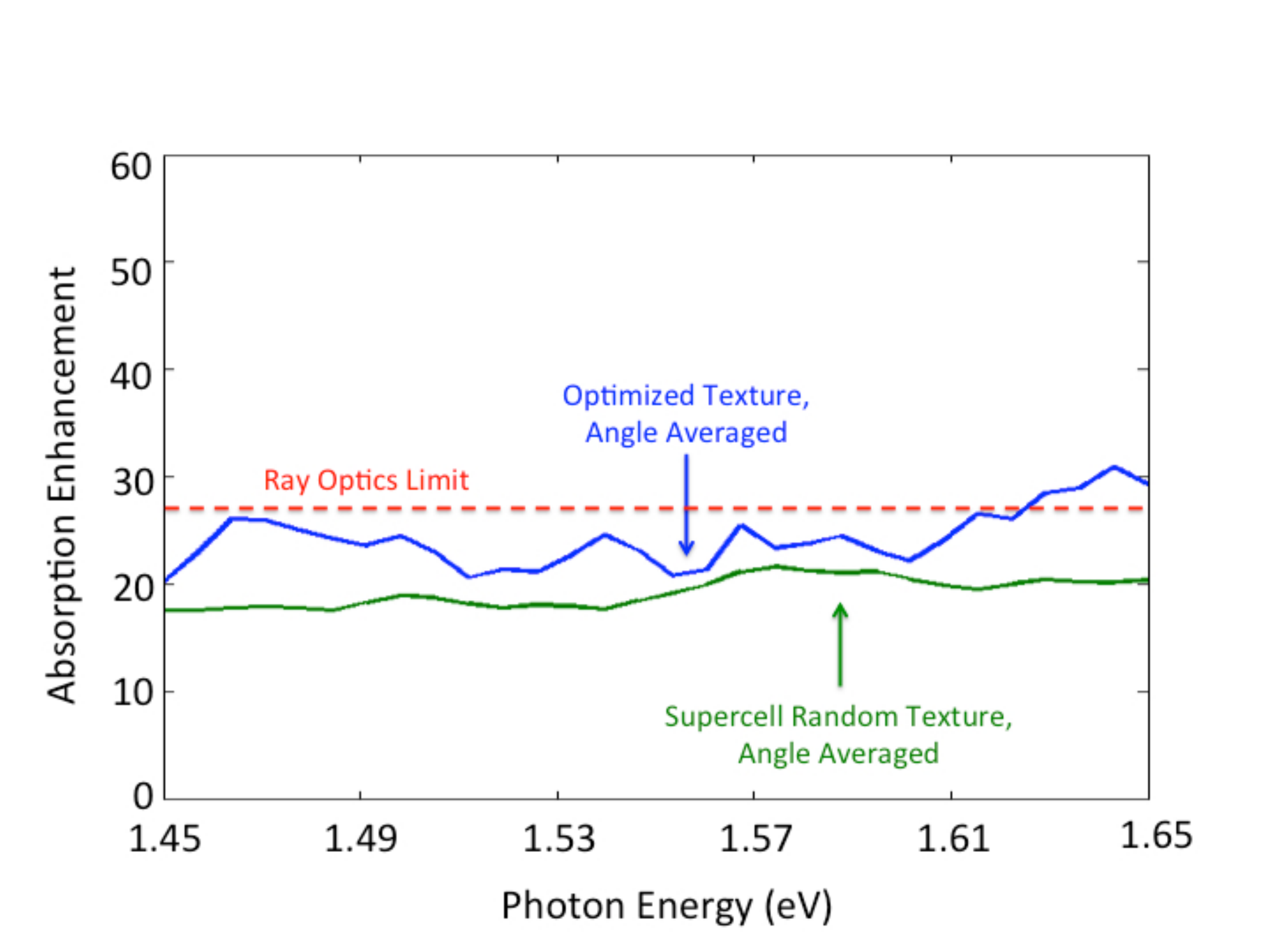}
\caption{The absorption enhancement as a function of frequency, angle averaged, for the optimized texture from Fig.~\ref{6broadOpt}(b) (blue) compared with the median random texture with 2300 nm periodicity from Fig.~\ref{17RandSuperCell} (green).}
\label{19SupercellCompareAngle}
\end{center}
\end{figure}

	We also check the performance of a completely random texture (i.e. a texture with infinite periodicity). We randomly generated Fourier coefficients to the 5th order for a periodicity that is $10\lambda_n = 3.5 = 2300$ nm, with a total texture amplitude between $\Delta h =$ 223 nm and 233 nm. Our large periodicity approximates a texture with infinite periodicity (the supercell approach). The resulting Figures of Merit for 11 different random supercell textures is shown in Fig.~\ref{16RandTex11}. The texture with the median Figure of Merit of $AE = 13$ is shown in Fig.~\ref{17RandSuperCell}; this median Figure of Merit is the same as for the random textures on a 710 nm periodicity. Fig.~\ref{18SupercellCompareNorm} compares the median supercell texture to the optimized texture at normal incidence. Additionally, the angle- and frequency-averaged performance (see Fig.~\ref{19SupercellCompareAngle}) of the optimized texture is 26\% better than the median random supercell. Our result that a periodic texture can perform better than a random one is in agreement with Ref.~\cite{battaglia_light_2012}. 

Our absorption enhancement factor ($AE$) results are summarized in Table 1.%~\ref{tab:FOM}.

\begin{table}[h]
\begin{center}
\label{tab:FOM} 
\caption{Absorption enhancement factor ($AE$) results.}
    \begin{tabular}{ | p{3cm} || p{2.7cm} | p{2.7cm} | p{2.7cm} | p{1.5cm} |}
    \hline
     & Best Optimized Texture & \raggedright Median Random Texture ($\Lambda =$ 710 nm) & \raggedright Median Random Supercell Texture ($\Lambda =$ 2300 nm) & Ray Optics Limit \\ \hline \hline %\noalign
\raggedright Figure of Merit (worst enhancement factor over frequency and polarization at normal incidence) & 29 & 13 & 13 & \\ \hline
\raggedright Angle- and Frequency-Averaged Enhancement Factor & 24 & 18 & 19 & 28 \\ %\hline
    \hline
    \end{tabular}
\end{center}
\end{table}

	In Table 1, the absorption enhancement factors are relative to a finite 1.6\% single pass absorption. To compare these results with the $4n^2$ ray-optics absorption enhancement limit, we need to account for the finite absorption in our structure. This can be done by using Eqn.~\ref{eq:rayOpticsA}, which can be written more generally as:  
\begin{equation}\label{eq:rayOpticsAMod} A = \frac{\alpha d}{\alpha d + \frac{1}{4 n^2}}= \frac{\alpha d}{\alpha d + \frac{1}{E}},\end{equation} where $E$ is the limiting enhancement factor when the single pass absorption is very weak ($\alpha d < 1.6\%$). $E$ represents the highest possible enhancement factor, which should be compared to the ideal $E = 4n^2 \approx 50$ case. For our optimized case of $AE = 24$ at $\alpha d = 0.016$, $E = 39$.

\section*{Conclusions}

	In the ray optics regime, random structures are optimal for achieving absorption enhancement \cite{yablonovitch_statistical_1982}. In the sub-wavelength regime, it appears that computationally optimized surface textures perform better than randomly generated ones. So far we have discovered a broad optimum, with many textures achieving similar figures of merit. We have shown that our optimized structures perform about $1.3\times$ better than randomly generated structures for angle- and frequency-averaged absorption. We report an angle- and frequency-averaged absorption enhancement factor in the weakly absorbing limit of $E = 39$, for a texture on a high index material of sub-wavelength thickness. This enhancement is $\approx 80\%$ of the ray optics limit $E = 4n^2 \approx 50$. 
	
	Though we do not prove a fundamental limit, the absorption enhancement factor arising from these optimizations is less than the ray-optics limit. It should be noted that for practical purposes, meeting or exceeding the ray-optics limit in the sub-wavelength might be unnecessary. For example, starting from a 1 $\mu$m film thickness, which for some materials makes a good solar cell even without light-trapping, an enhancement factor $E = 50$ would permit a reduced film thickness of 20 nm which is almost too thin for manufacturing purposes. A more reasonable 100 nm solar cell thickness requires an enhancement $E \approx 10$, which is easily achieved.  
	
	In evaluating the performance of a solar cell texture for light trapping, it is important to take into account both the average performance, as well as the worst performance over frequency. A texture with a few resonant peaks may yield a high average performance in theory, but when applied to a real material, the resonant peaks will saturate at 100\% absorption, and the total photons absorbed will be low. This electromagnetic optimization procedure obtains both a broadband absorption spectrum and a high average absorption. 
	
	Our practical goal in designing a solar cell texture is to achieve complete light absorption in the thinnest possible layer, with a manufacturable texture. This optimization procedure can be used as a tool in designing textures for real materials. In addition, we can add manufacturing constraints to the problem. For example, in these calculations, we permitted the minimum film thickness to drop as low as 1 nm, but we could constrain to a more realistic minimum thickness of 50 nm. In our procedure, we also could have fallen into local optima and never found the global optimum. Further work requires us to investigate the question of whether we have really converged.

\section*{Acknowledgements}

This work was supported by the DOE ÒLight-Material Interactions in Energy ConversionÓ Energy Frontier Research Center under grant DE-SC0001293. V. G. acknowledges support from the Department of Energy Office of Science Graduate Fellowship Program (DOE SCGF), made possible in part by the American Recovery and Reinvestment Act of 2009, administered by ORISE-ORAU under contract no. DE-AC05-06OR23100. This research used resources of the National Energy Research Scientific Computing Center, which is supported by the Office of Science of the U.S. Department of Energy under Contract No. DE-AC02-05CH11231.

\bibliographystyle{ieeetr}
\bibliography{light-trapping-paper}

\end{document}